\documentclass[11pt,a4paper]{article} 
\usepackage{bm,amsmath,amssymb,slashed,graphicx,%
            enumerate,alltt,xspace,multirow,xcolor,mathrsfs,url}
\begin{document}

\begin{titlepage}

\begin{flushright}
CERN-TH-2017-266\\
\noindent \today

\end{flushright}

\vskip1cm
\begin{center}
{\fontsize{15}{19}\bf\boldmath
The Top Mass in Hadronic Collisions}
\end{center}

\vspace{0.8cm}
\begin{center}
{\sc P.~Nason} \\
\vspace{0.3cm}
Theoretical Physics Department, CERN, Geneva, Switzerland, and\\ INFN, Sezione di Milano Bicocca, Milano, Italy.
\vspace{0.3cm}
\\[0.3cm]
\end{center}

\vspace{0.7cm}
\begin{abstract}
\noindent \normalsize
I discuss theoretical issues related to the top mass measurements
in hadronic collisions.
\end{abstract}

\vspace{2cm}

\begin{center}
Contribution to the volume \\ \vspace{0.3cm} \emph{“From My Vast Repertoire”} -- \emph{The Legacy of Guido Altarelli.}
\end{center}
\end{titlepage}

\newcommand\PNmcMSbar{\ensuremath{\overline{\rm MS}}}
\newcommand\PNmcMSR{\ensuremath{{\rm MSr}}}

\newcommand\PNmccomment[1]{{\bf Comment: #1}}

\newcommand\PNmcPOWHEG{{\tt POWHEG}}
\newcommand\PNmcMCatNLO{{\tt MC@NLO}}
\newcommand\PNmcpythiaEight{{\tt Pythia8}}
\newcommand\PNmcherwigSeven{{\tt Herwig7}}
\newcommand\PNmcas{\ensuremath{\alpha_s}}
\newcommand\PNmcasymplim{a}

\section{Introduction}

There is something strange about the current status of top mass
measurements at hadron colliders.  Regarding the most precise
measurements, performed by kinematic reconstruction of the full $t\bar{t}$ event,
and commonly referred to as ``Standard
Measurements'', often it is not stated clearly what is measured, in
contrast with the very small error that is proudly quoted by the
experimental collaborations.  In some
circumstances, it is claimed that what is measured is a ``Monte
Carlo'' mass, i.e. just a parameter in the generator used to perform
the analysis. But yet extensive variations of Monte Carlo parameters
are performed in order to estimate errors, in a clear effort to reach
something more fundamental.  Other measurements, for example from the
total cross section, or from some kinematic distributions, are
presented collectively as ``Pole Mass Measurement'',
that is (in the context of perturbation theory)
a theoretically well defined parameter.
No attempt is
made to combine these pole mass measurements with
the direct measurements.\footnote{I have heard a
presentation where the speaker has jokingly introduced
the subjects saying that the top is the only Standard Model particle
with more than one mass.}

I remember having a discussion on this issue with Guido Altarelli, many
years ago, at an Italian conference on LHC physics. Much time has gone by,
and I do not remember the exact year and conference.  A recurring
argument was circulating among theoretical physicists,
stating that top mass measurements at hadron colliders were
just extracting the Monte Carlo mass, a parameter that did not have
a well defined  field theoretical
definition. The argument went as follows: the top mass is extracted
by fitting Monte Carlo templates to measured distributions;
Monte Carlo generators
have just leading order accuracy; different mass renormalization schemes
cannot be distinguished at leading order, since renormalization enters
at least at one loop; hence the Monte Carlo mass is not in a definite
mass scheme. One-line arguments like this are quite catchy, and
tend to spread very quickly, since they do not require any specialized
knowledge to be understood.
I strongly disagreed with it. The measurements
were clearly aiming at observables that were strongly correlated to
the mass of the system of decay products, i.e. were aiming at
the pole mass. And the error associated with the LO accuracy of the
Monte Carlo were estimated by several methods, typically by varying
parameters in the Monte Carlo, and by comparing its output to more
precise next-to-leading order calculations.
I expressed my disappointment about the diffusion of this ``Monte Carlo mass''
concept to Guido Altarelli.
I told him that the argument was just used
to scare the experimentalists with theoretical concept that they were not
very familiar with, which was in fact an aspect of the issue
that I was finding most irritating. Guido's answer was that, rather than scaring them, the argument
should have made them think.
I could not reply to that, since for an instant I thought that perhaps
he was subtly implying that I myself should have done more thinking.
This was quite typical of my exchanges with Guido. They
were never free of some sort of friction.
And that was certainly not the only occasion when I departed from a discussion with
him with a slightly sore feeling.
I must admit that I was never great at verbally arguing
about physics issue, while Guido was clearly excellent at it. Yet, I was always eager to discuss with
him. I had the perception (that I believe was shared by many others)
that in order to settle an issue you had to convince him,
since he was \emph{the} uncontested authority in our field, and with the good reasons that we all know.
And discussing with him always helped me to acquire a more detached and focused
view of the problems.
It is unfortunate that this discussion stopped at that point, since I
didn't worry about this issue any more until recent time.

In this work I express my personal views on this subject, that at the moment
is undeniably very controversial.

In the first section I will illustrate how and why the
current status of the theoretical interpretation of top mass
measurements has been reached. I will not try to fully reconstruct the
historical development of all related arguments, and thus I will certainly omit
quoting more than one important paper on this topic. I will rather
focus on the arguments and papers that have had more influence in
determining the current status. The message that I would like to
convey is that, rather than finding or denying some relation of the
Monte Carlo mass parameter to some well defined field theoretical
parameters, we should find ways to estimate the error on the
extraction of a theoretically well defined mass parameters when using
a Monte Carlo generator.

In the second part of this work I will illustrate what I believe
has been a small but useful progress, i.e. an improved understanding of the relation
of the pole mass to the \PNmcMSbar{} mass. It is a self-contained
subject where analytic calculations, not requiring modeling of
non-perturbative and hadronization effects, can lead to a partial
progress in our understanding of the top mass issues. I will expose this topic
avoiding technicality as much as possible, yet discussing what
I believe are its most relevant aspects.

Needless to say, I am trying to convince Guido.
\section{The top mass problem}
The top quark mass is a
key parameter of the Standard Model. Its large size, of the
order of the Electro-Weak scale, is associated with a
Yukawa coupling of order~1, that gives important contributions,
via radiative corrections, to Standard Model observables.
After the Higgs boson
discovery and the accurate measurement of its mass, the allowed values of
the $W$-boson and top-quark masses have become strongly correlated,
so that an accurate determination of both would lead to a SM
test of considerable precision~\cite{Patrignani:2016xqp,
  Baak:2014ora}. At present there is some tension, at the
$1.6\,\sigma$ level, between the indirect top-mass determination from
electroweak precision data~($177\pm 2.1$~GeV) and the combination of
direct measurements at the Tevatron and at the LHC~\cite{ATLAS:2014wva},
that yields~$173.34\pm 0.76$~GeV. More recent
determinations~\cite{Aaboud:2016igd, Khachatryan:2015hba, CMS:2017eoz,ATLAS:2017lqh}
favour an even lower value, close to $172.5$~GeV.

The top mass value is also critical in the issue of vacuum stability
in the Standard Model~\cite{Degrassi:2012ry, Buttazzo:2013uya,Andreassen:2017rzq}.
At high scales, the Higgs quartic coupling $\lambda$ evolves to
increasingly small values as $m_t$ grows, and above $m_t=171$~GeV,
i.e.~very close to the present world average, $\lambda$ becomes
negative at the Planck scale, rendering the electroweak vacuum
meta-stable, while for $m_t>176$~GeV the electroweak vacuum becomes
unstable. The only conclusion that can be drawn from this result
is that the current value of the Top and Higgs masses are such
that no indication of new physics at any scale can be inferred
by imposing the stability or metastability of the vacuum. On the other
hand, the fact that the Higgs quartic coupling nearly vanishes
at the Planck scale may have some deep meaning that we are now unable to unveil.

The abundant production of top pairs at the Large Hadron Collider~(LHC)
provides an opportunity for accurate top mass measurements,
that are generically performed by fitting $m_t$-dependent
kinematic distributions to Monte Carlo predictions. The most precise ones
rely upon the full or partial reconstruction of the system of the $t$ and
$\bar{t}$ decay
products.  The CMS measurement of Ref.~\cite{Khachatryan:2015hba}, yielding
the value $m_t=172.44 \pm 0.13 ~{\rm (stat)} \pm 0.47 ~{\rm (syst)}$~GeV,
falls into this broad category.\footnote{A similar measurement performed at
13~TeV~\cite{CMS-PAS-TOP-17-007}
yields a value of $172.25 \pm 0.08(stat+JSF) \pm 0.62(syst)$~GeV.}

In contrast with the increasing experimental precision with which the
Top mass is measured at the LHC, the theoretical interpretation of the
measurements seems to be in a questionable state. The so called
``direct measurements'', i.e. those that rely upon the
reconstruction of the kinematics of the $t\bar{t}$ system,
have been heavily criticized by some theorists as not possessing a clear
relation between the extracted mass and a well defined quantum field
theory parameter of the underlying theory.

As a consequence of that, in many
experimental papers and talks, it is preferred not to qualify the
measured mass parameter with a precise field-theoretical attribute,
such as the ``pole mass'' or the ``\PNmcMSbar{} mass'', and, at times,
even to qualify it as a ``Monte Carlo'' mass
\cite{Patrignani:2016xqp,ATLAS:2014wva,CMS:2015uoa}.  This has led to
the paradoxical situation that the most precise measurement available
is not receiving the attention that it deserves from the theoretical
community, and the interpretation of the measurement is left in a
``limbo'', with the hope that some theoretical work may clarify
it in the future.  At the same time, several theorists have
suggested observables that allegedly overcome the problems of the direct
measurements. Although all these suggestions are valuable, since they
can provide alternative determinations and consistency checks, none of
them seem to possibly lead to an accuracy comparable with the direct
determination. Furthermore, these alternative measurements are grouped
in a different category by the experiments, that do not attempt
to combine them with the standard measurements.

The aspect of the top mass measurement that is mostly puzzling the theoretical
community, is that the top mass cannot be defined in terms of
the mass of the system of its decay
products. Since top is a coloured object, no final-state hadronic system can
be unambiguously associated with it. On the other hand, the top
mass bears certainly some relation to the mass of the system of objects
arising from its decay (i.e.~leptons, neutrinos and hadronic jets). The mass
distribution of this system can be computed, and the top mass
enters as a parameter in this computation. This is in fact the case
for many parameters that are measured in high energy physics experiment.
In the case of top, however, the computation in question is performed
by a parton shower generator. Thus the idea that what is measured is
a ``Monte Carlo'' mass. This idea has been expressed by several authors,
but not always in the same sense: one line of thinking has to do with the
perturbative accuracy of Monte Carlo generators, while others
worry about the non-perturbative effects that they model.
As far as the ``perturbative accuracy'' is concerned, the argument
is essentially as follows: since shower Monte Carlo are only accurate
at leading order, they cannot possibly distinguish mass parameter definitions
that differ only at next-to-leading order, like the pole mass and the
\PNmcMSbar{} mass, whose difference amounts to several GeV's.

In the following I will examine more carefully the arguments and the studies
that have led to the current situation. I will show that the
most disturbing aspect of the ``Monte Carlo Mass'' concept is its ambiguous
role. In other words, it has acquired a different significance according to different
authors, up to the point where it is used to support conflicting
points of view.
\subsection{The ``Perturbative'' argument}
Fixed order theoretical calculations of (infrared safe) final state distributions
can be catalogued according to their perturbative accuracy as Leading Order (LO),
next-to-leading order (NLO) and so on. The top mass parameter in a theoretical
calculation must be defined within a given renormalization scheme, since
(divergent) perturbative corrections arise order by order in perturbation theory.
For the top mass parameter, one such scheme is the pole mass scheme,
that prescribes to subtract the
divergent mass corrections in such a way that the pole in the quark propagator remains
fixed order by order in perturbation
theory. Alternatively, the \PNmcMSbar{} scheme prescribes to subtract the pure $1/\epsilon$
pole in the divergent mass correction. Thus, in the \PNmcMSbar{} scheme the position of
the pole in the top propagator receives corrections at all orders in perturbation theory.
The two definitions lead to perturbatively equivalent theories, in the sense that there
is a perturbative expression of the pole mass in terms of the \PNmcMSbar{} mass that allows one
to translate a physical result in the pole mass scheme into the corresponding physical
result in the \PNmcMSbar{} scheme. We notice that the \PNmcMSbar{} mass begins to differ from the
pole mass at first order in perturbation theory, that contributes about 7.5~GeV to the
total difference.

We report below the relation between the top pole mass $m_p$ and \PNmcMSbar{} mass $m$
up to the four loops level~\cite{Marquard:2015qpa},
\begin{equation}\label{eq:MSB_Pole}
  m_{p} = m + \underbrace{7.557}_{\rm NLO} + \underbrace{1.617}_{\rm NNLO}
    + \underbrace{0.501}_ {\rm N^3LO} + \underbrace{0.195 \pm 0.005}_{\rm N^4LO} {\rm GeV}.
\end{equation}
for $m=163.643\,$GeV and $\alpha_s^{(6)}(m)=0.1088$, in order to
show the typical numbers that come into play.
Thus, the ``perturbative argument'', if pushed to its extreme,
would lead us to conclude that the relation of the extracted mass parameter to some
physically well defined one has an ambiguity of the order of  10~GeV.

The earliest written record of this argument that I found is
in ref.~\cite{Alioli:2013mxa},
where it is stated quite clearly that ``the top-quark mass derived from the
kinematical reconstruction does not correspond to a well defined
renormalization scheme leading to a theoretical uncertainty in its interpretation.
Nevertheless it is usually interpreted as the top-quark pole mass''.
In the same work
it was proposed to extract the top mass by comparing kinematic observables in
$t{\bar t}$ events involving one additional jet. Such process can be computed at
NLO level, and so it must use a well defined mass renormalization scheme.\footnote{
It should be recalled, however, that
the difference between the pole and the \PNmcMSbar{} mass
receives contributions beyond the NLO
order that amount to more than 2~GeV. Thus, one may also argue that in an NLO
calculation the pole mass parameter used there is related to a fundamental
parameter of the theory with an uncertainty of at least 2~GeV.}

This argument has also been used to support mass measurement
based upon the $t\bar{t}$ cross section,
that is now computed at NNLO order~\cite{Czakon:2013goa}.

Sometimes, all the mass measurement techniques that refer
to a fixed order calculation beyond the Leading Order are collected
together and presented as ``measurements of the top pole mass'',
in contrast with direct measurement where
the mass scheme is never specified (see
\url{https://twiki.cern.ch/twiki/pub/CMSPublic/PhysicsResultsTOP/pole_mtop.pdf}).

\subsection{The ``Non Perturbative'' argument}
There are claims in the literature that the difference between the top pole mass and the mass
extracted in direct methods arises due to non-perturbative effects, that
are only modelled in a shower Monte Carlo. This claim is certainly
true, although the difficult bit in it is to quantify this difference.
Again, this difference is used to motivate a ``Monte Carlo mass''
concept, that cannot be related to a well defined theoretical parameter
because of these non-perturbative effects.
It does not have the same meaning as the ``Perturbative'' difference
discussed previously. But also if we restrict ourselves
to this ``Non Perturbative'' meaning, different authors interpret it
differently.

In ref.~\cite{Kawabataa:2014osa}, the direct measurements
are criticized on the basis that they reconstruct the top
mass making use of jets, and jet reconstruction is affected
by hadronization effects that are modelled by the shower Monte Carlo.
In contrast, they propose using observables that only depend upon the
lepton kinematics and that are insensitive to production dynamics,
and thus should not (in their opinion) be affected
by hadronization effects.

Other kind of lepton observables have also been advocated in ref.~\cite{Frixione:2014ala},
There, however, an attempt is made to single out those observable that seem
less affected by shower and hadronization effects, on the basis
of Monte Carlo simulation study. In other words, the authors rightfully
recognize that also leptonic observables may be subject to
hadronization corrections, and try to quantify these effects by Monte Carlo studies.

A considerably more elaborated argument was first put forward in
ref.~\cite{Hoang:2008xm}, where it is stated quite explicitly that
``it is not $m_t^{\rm pole}$
that is being measured by the Tevatron analyses''.
The argument goes through several steps, that I can only summarize here.
First of all, it is argued that the pole mass scheme is a poor choice,
because of the presence of an ambiguity of order $\Lambda_{\rm QCD}$
associated with the mass infrared renormalon (that I will discuss
later in this work), and that the so called \PNmcMSR{} scheme
avoids this problem. The  \PNmcMSR{} mass is a mass parameter that depends
upon a scale $R$. It is defined using the same self-energy diagrams that
contribute to the pole mass, but their contribution is limited to loop
momenta of order larger than $R$. Thus, as $R$ becomes small, the  \PNmcMSR{} mass
approaches the pole mass. It is argued in~\cite{Hoang:2008xm} that an  \PNmcMSR{} mass,
with $R$ taken of the order of 1~GeV, should be used as mass parameter.

The argument is further developed
by considering top production at high momentum in $e^+e^-$ collisions
at energies much larger than the top mass. It is argued that the hemisphere
masses (defined by dividing the event with a plane orthogonal to the trust
axis) are calculable using SCET techniques, up to the inclusion
of power suppressed effects of order $\Lambda{}$, provided the
mass parameter is the \PNmcMSR{} mass. From this framework, implications
for the standard measurements at the Tevatron are drawn. However, they suffer for the fact
that in standard measurement the top is not ultrarelativistic, and the SCET
factorization cannot be applied.

It must be stressed that in ref.~\cite{Hoang:2008xm} it is recognized
that the Monte Carlo mass parameter should be quite close to the
pole mass. More specifically, it is claimed that it should
be close to the \PNmcMSR{} mass evaluated at a small scale, that
in turn is close to the pole mass.
This point of view is very different from the ``perturbative''
one, that leads to differences of order $\alpha_s m_t$
between the pole and the Monte Carlo mass.

There are several followups of the work in ref.~\cite{Hoang:2008xm}.
In  ref.~\cite{Hoang:2014oea}, it was argued that, in general,
an additional uncertainty of 1~GeV should be accounted for
in top mass measurements at hadron colliders. It is also
argued that the top pole mass in general
cannot be determined with a precision better than
1~GeV because of the mass renormalon problem.
Although it is stated somewhere in this paper that 1~GeV ``is the energy
value I use in this talk for what theorists call hadronization scale'',
this value, quoted in the abstract and in many places in the paper,
has been echoed as is in several other publications and talks, to be
taken as a serious hard limit on the precision that can be achieved
in the measurement of the top mass by direct methods.
In subsequent works~\cite{Butenschoen:2016lpz},
Hoang and collaborators attempt to quantify
a relation between the so called Monte Carlo mass and an \PNmcMSR{} mass
evaluated at 1~GeV. Essentially, always in the framework of highly boosted
tops, they compute the jet mass using both an NNLL SCET calculation,
and the Pythia8~\cite{Sjostrand:2014zea} shower Monte Carlo. They absorb the difference in the two
results by a shift in the Pythia8 mass parameter with respect to the
\PNmcMSR{} mass used in SCET, and argue that
this shift is the difference between the Monte Carlo and the \PNmcMSR{} mass.
Indeed, they find that the Monte Carlo mass exceeds the \PNmcMSR{} mass
by about 200~MeV with an uncertainty of about the same size. On the other hand,
also the pole mass exceeds the \PNmcMSR{} mass by a similar amount.
Whatever method one adopts to resum the divergent series that gives
the pole mass in terms of the \PNmcMSR{} mass, it seems that the difference
between the pole mass and the Monte Carlo mass should
be significantly less than 1~GeV.

We also observe that in \cite{Butenschoen:2016lpz},
effects that are accounted for in the SCET results (that is supposed to be NNLL accurate)
and are not present in Pythia8  (that certainly cannot claim this accuracy) are absorbed
into a shift in the mass. It thus seems
that the mass shift they find, rather than having a universal character, should
be dependent upon the chosen observable.

\subsection{Summary of the issue}
We have seen at least three arguments that are used to
criticize the standard top mass measurements:
\begin{enumerate}
\item The perturbative argument (i.e., MC are only LO accurate)
\item A non perturbative argument: jets have non-perturbative
  corrections, thus we should avoid top mass observables that involve jets,
  (i.e., use leptonic observables).
\item A deeper non perturbative argument: since MC's implement ad hoc models
  of non-perturbative effects, we should not rely upon them, since they
  imply an irreducible non-perturbative error. Rather,
  we should resort to methods that allow to treat non-perturbative corrections
  from first principles.
\end{enumerate}
These arguments have different meaning, and in particular the first one is independent
from the remaining two, since it deals with perturbative effects, while the
other two address non-perturbative issues.
Yet, for example, in ref.~\cite{Kieseler:2015jzh}
it is stated that the translation from the Monte Carlo mass
to a short distance mass can only be estimated within a 1~GeV
accuracy, quoting in particular~\cite{Hoang:2008xm},\footnote{
A few months later, this 1~GeV accuracy would have gone down to 200~MeV
in ref.~\cite{Butenschoen:2016lpz}.} and this statement
is used to advocate
that it is preferable to extract the
top mass by considering observables that can be calculated
in QCD at least at NLO. It is unclear why NLO precision should
get rid of non-perturbative effects.

Regarding the first of the three arguments listed above, it should
be reminded that one cannot simply say that a Monte Carlo has only LO
accuracy. The accuracy of a Monte Carlo depends upon the observable,
and there is no simple statement that can qualify it for all
observables.

The production and decay dynamics of a coloured resonance in the
narrow-width limit factorize in perturbation theory, and become
independent, as a consequence of the fact that the resonance can
propagate a long time before decaying. Thus, in the narrow width limit
and in perturbation theory, there is an unambiguous definition of the
system of the resonance decay products, whose mass coincides with the
resonance pole mass to all orders in perturbation theory if the pole
mass scheme is used. The factorization feature of perturbation theory
is also implemented in Shower Monte Carlo generators, where radiation
in production and decay are developed independently, and preserve the
mass of the decaying resonance. Thus, up to non-perturbative effects
and soft radiation of energy comparable to the resonance width (that
can violate factorization because of interference between production
and decay) the resonance mass appearing in the Monte Carlo can be
identified with the pole mass.

One could still argue that the accuracy of the Shower Monte Carlo
generator does play a role in the mass determination. Typically, in
top production, the radiation generated at the production stage can
enter the $b$-jet cone, or radiation from the $b$-quark can escape the
jet cone, thus altering the mass of the reconstructed decay
system. These effects can lead to an error on the extracted mass. This
error, however, can be reduced by improving the perturbative accuracy
of the Monte-Carlo, or by tuning the Monte Carlo in such a way that it
better describes radiation in production and the structure of the
$b$-jet. In other words, these errors are modeling errors, and their
impact can be estimated (and in fact is estimated by the experiments)
by varying suitable parameters in the shower Monte Carlo.  But none of
these errors is associated to problems in the perturbative definition
of the top mass.

Fixed order calculations of kinematical distributions involving top
quarks are also subject to soft radiation and non-perturbative
effects, and thus cannot be considered as privileged observables for
pole mass determinations, to be presented in isolation from those
obtained with direct measurements. If the former can be considered
pole mass determinations, the same holds for the latter.  Notice that
this also holds for total cross section. Even if one is willing to
believe that the total cross section has no linear power corrections
(i.e. a correction suppressed by a single power of a hadronic scale
divided by the top mass), one should remember that total cross section
measurements requires extrapolations outside of a fiducial region, and
fiducial regions have cuts that can introduce linear power
corrections.  Associated non-perturbative effects must thus be
estimated also in this case.

We can also add that the accuracy of the Monte Carlo generators
currently used in standard measurements make use of \PNmcPOWHEG{} or
\PNmcMCatNLO{} generators, including NLO corrections in production.
Most modern Monte Carlos, like
\PNmcpythiaEight{}~\cite{Sjostrand:2014zea} and
\PNmcherwigSeven{}~\cite{Bellm:2015jjp}, include matrix element
corrections to top decay, that makes them essentially (up to an
irrelevant normalization factor) NLO accurate in
decay. \PNmcherwigSeven{} also implements its own \PNmcPOWHEG{}
implementation of NLO corrections to top decay.  Furthermore, there
are recent NLO+PS generators that implement top production and decay
including finite width, non-resonant effects and interference of
radiation in production and decay~\cite{Jezo:2016ujg}.

The second argument is based upon the general feeling that
leptonic observables should not be affected by hadronization effects.
It is easy, with a simple example, to convince ourselves that this may
not be the case.
Consider the decay of the top into a $W$ and a $b$ jet.
The emission of an extra soft pion at large angle with respect to the $b$ jet
can represent a typical non-perturbative effect.
Such emission would always subtract an energy of the order of few hundred MeV's
from the $W$, thus reducing also the lepton energy by few hundred MeV's.
Thus, it is likely that linear power corrections are also present
in leptonic observables, and a
corresponding uncertainty should be estimated and associated with them.

The third item in the list should be regarded more carefully. It is
certainly undeniable that shower Monte Carlo do not implement a solid
theory of the leading power suppressed effects. It would certainly be
desirable to have such a theory, or to perform top mass measurement
in frameworks where such a theory is available. This seems to be the case
for top mass measurements performed by a threshold scan of the $e^+e^-$
cross section (see \cite{Beneke:2017rdn} and references therein),
or for the recently proposed mass measurement
based upon the shape of the $\gamma\gamma$ mass spectrum at the LHC~\cite{Kawabata:2016aya}.
While the first possibility is conditioned by the future developments
of high energy physics experiments, the second one seems to be limited
by statistics even at the high luminosity LHC phase.

We have seen that there is a sequel of papers
that advocate the use of highly boosted top-quark jets,
claiming that they can
be computed up to the leading power suppressed correction.
These approaches require validation against data,\footnote{We should
  remember that this is always the case in perturbative QCD, that is a
  consistent framework, but is based upon some unproven assumptions.}
full scrutiny and criticism by the theoretical community, and, most
important, should be demonstrated to be practically useful, i.e. they
should lead to top mass measurements with an an error
smaller than the leading power suppressed effects that they claim to
model correctly.  At the moment, these conditions do not seem to be
fulfilled.

So, the question remains about what to do with top mass measurements by
direct methods at hadron colliders. It is obvious that we have an
error due to non-perturbative effects that should be estimated,
but it is not enough to state that it is an error of the order of 1~GeV.
We should understand whether it is 1 or more, 0.5, 0.2 or less GeV's.
Some authors, typically those advocating the use of boosted
tops, have argued in the past for uncertainties
of non-perturbative origin, of magnitude near 1~GeV, affecting
the relation of the Monte Carlo mass, and also of the pole mass,
to some short distance mass. As we have seen, however, the 1~GeV
figure is scarcely motivated as a hard limit. One also gets the impression
that these ``near a GeV'' uncertainty keep decreasing with time, as more
thorough analyses are performed, leaving us wondering why such a large
value was ever quoted in a first place.

\subsection{What to do}
We have seen that the theoretical problems related
to the determination of the top mass at hadron
colliders boil down to the problem of quantifying
how power suppressed corrections affect the measurement.
More specifically, it has to do with the uncertainties
associated with how Monte Carlos implement
power suppressed effects, and how they match them to
perturbative effects. These Monte Carlo uncertainties
have been translated by some authors into the concept
of a Monte Carlo mass.

Translating a Monte Carlo inaccuracy into a Monte
Carlo mass concept has also led to the equation:
the Monte Carlo implements
perturbation theory at leading order, thus its mass
parameter cannot be associated with a well defined mass
scheme. In this second case, it is quite clear that the
right question to ask should have been: what is the
error on the measurement due to the fact that
the Monte Carlo is only accurate at leading order.
This question has in fact been asked by the experimentalists,
that have studied the errors associated with the MC
perturbative inaccuracies with several techniques,
and by theorists, that have developed improvements
in the generators to promote the accuracy of the Monte
Carlo to the NLO level.

The question now is whether uncertainties due to
power suppressed effects can be reliably estimated
in similar ways, i.e. by varying Monte Carlo
parameters, or by including plausible alternatives
in the model of hadron formation that they implement.
This is currently done by the experimental collaborations and by
Monte Carlo developers.\footnote{As an example,
  the increase in the error in the 13~TeV top mass
  measurement of CMS reported in \cite{CMS-PAS-TOP-17-007}
  relative to the previous 8~TeV measurement
  is due to the introduction of alternative models
  of colour reconnection in {\tt Pythia8} (see \cite{Argyropoulos:2014zoa,Christiansen:2015yqa}).}
Yet, with these methods
the doubt remains that we may be missing something,
and investigations aiming at a better understanding
of power suppressed effects, and their eventual
interplay with renormalons, would be welcome.
At the moment, however, there is no sound argument
that suggests that the Monte Carlo mass is related
to some well-defined mass parameter via a systematic
shift. It has been argued by some authors that, since
shower Monte Carlo have a 1~GeV cut off on soft
radiation,\footnote{%
  We note again the loose use of the 1~GeV figure.
  In Pythia8, such cutoff is by default 500~MeV.}
their mass parameter should be identify
with an \PNmcMSR{} mass at a scale $R=1$~GeV, rather
than a pole mass, which yields a difference of few hundred
MeV's. It should be recalled, however, that Monte Carlos
implement soft radiation by ensuring that virtual effects
cancel completely the real emission corrections in
inclusive quantities, via a mechanism known as ``shower
unitarity''. This mechanism implements in practice the
cancellation of soft singularities. However,
also finite, non-singular soft effects are cancelled out
in this way, and we should remember that self energy mass
corrections are non-singular. Thus, the ``Shower cut off''
argument does not track down Monte Carlo
effects that can convert a pole mass to an \PNmcMSR{} one.

Notice that it is also easy to set up simple and catchy arguments to
prove that the Monte Carlo mass parameter \emph{is} the top pole mass.
One could argue, for instance, that, since
Monte Carlos implement a top propagator with a
complex pole at a fixed position, that pole must be at
the (complex) top pole mass, since the use of a short
distance mass would yield a pole position that is
blurred by the mass renormalon effects. Needless
to say, also this argument is not very convincing, and
it is clear that these issues require more serious
thinking. However, it doesn't seem appropriate to tell
the experimental community that until such thinking
is done they are not allowed to tell what they are actually
measuring.
Rather than telling the experimentalists to wait
for new theoretical developments to interpret their
results, it would be far more constructive to point
out to them methods to estimate the uncertainties associated
with the limitations of the generator they use.
For example, if the lack of radiation below the Monte Carlo cutoff
scale is a concern, one should recall that this cutoff scale can be
changed. One could setup a different tune of the Monte Carlo, using
a different cut-off scale, and examine whether it leads to relevant
differences in the extracted top mass. I would also like to stress
that this work should not be necessarily done by the experimentalists.
It can be done by theorists as well, using simplified version
of top mass observables and of detector effects.

\section{The renormalon problem}
If the renormalon ambiguity on the top pole mass was as large as 1~GeV,
it would make no sense to
measure the top pole mass with the experimental precision that
is quoted today. Fortunately, it turns out that this ambiguity is in fact much
smaller. In refs.~\cite{Beneke:2016cbu} and~\cite{Hoang:2017btd} the relationship
of the pole mass to the \PNmcMSbar{} mass for the top quark is studied. They find
compatible results for the central values, while for the renormalon ambiguity
they quotes values of 110~MeV and 250~MeV respectively,
both safely below currently quoted systematic errors.\footnote{In fact,
  values in this range were obtained much earlier in refs.~\cite{Pineda:2001zq,Bali:2013pla},
  mostly in a bottom physics context, but since the renormalon ambiguity does
  not depend upon the heavy quark mass, they also apply to top.}
  
In the rest of this article I will briefly review the mass renormalon problem.
I will also try to clarify why the quoted errors are so different in the two publications.

\subsection{A simplified description of the renormalon problem}
The quark mass parameter $m$ that we introduce in our Lagrangian does not
necessarily coincide with the quark pole mass, which is
the position of the pole in the quark propagator. More precisely
it coincides with it only at zeroth order
in perturbation theory. At higher orders the mass parameter requires
renormalization, but this can be carried out without ever referring to
the pole position, like for example when using the \PNmcMSbar{} prescription.

In the following I will denote with $m$ the mass parameter renormalized
in the \PNmcMSbar{} scheme.\footnote{It is common practice to
  use the term ``\PNmcMSbar{} mass'' (that we call $m$ in this paper)
  to denote the running \PNmcMSbar{} mass, $m(\mu)$,
  that depends upon a scale $\mu$, evaluated self-consistently at the scale $\mu=m(\mu)$.}
The ${\cal O} (\alpha_s)$ correction to the position of the pole in the
heavy quark propagator turns out to have a linear infrared sensitivity
to the scale of the momentum flowing in the loop,
i.e. to yield a contribution of the form
\begin{equation}
  \int_0^m {\rm d}l\, \alpha_s,
\end{equation} 
where $l$ is the scale of the loop momentum.
The upper cutoff arises since we have assumed that
ultraviolet divergences related to the large $l$ region have been subtracted by
renormalization. Here we assume that the renormalization scale is taken equal to $m$.
Thus, the pole position is given by an expression of the form
\begin{equation}
  m_{\rm P}=m+N \int_0^m {\rm d}l\; \alpha_s + {\cal O}(\alpha_s^2),
\end{equation}
where $N$ is a suitable normalization factor.

We now assume heuristically that we can obtain the higher order terms by simply
writing
\begin{equation} \label{PNmc-eq:poleMSsimple}
  m_{\rm P}=m+N\int_0^m {\rm d}l\, \alpha_s(l),
\end{equation}
with
\begin{equation}\label{PNmc-eq:alphaexpansion}
  \alpha_s(l) = \frac{1}{b_0 \ln\;l^2/\Lambda^2}=
  \frac{\alpha_s(m)}{1-\alpha_s(m)\,b_0
    \ln\,m^2/l^2}=\sum_{n=0}^\infty \alpha_s^{n+1}(m)\,b_0^n \ln^n\frac{m^2}{l^2}\,.
\end{equation}
Inserting (\ref{PNmc-eq:alphaexpansion}) in (\ref{PNmc-eq:poleMSsimple}),
and recalling that
\begin{equation}
  \int_0^m {\rm d}l\;\ln^n\frac{m^2}{l^2}=m\, 2^n\int_0^1 {\rm d}x\;\ln^n\frac{1}{x}=m\, 2^n n!
\end{equation}
we get
\begin{equation}
\label{PN-eq:selfenexp}
m_{\rm P}=
 m+N \sum_{n=0}^\infty m \left(2 b_0\right)^n \alpha_s^{n+1}(m)\,n!\,.
\end{equation}
The factorial growth of the coefficients of the perturbative expansion is
what is called ``renormalon'', where the name is suggested by the fact that it
arises due to the renormalization group evolution of the coupling constant.
It leads to a power expansion that has
zero radius of convergence. This is related to the fact that the integral in
eq.~(\ref{PNmc-eq:poleMSsimple})
runs over the Landau Pole, i.e. the divergence of $\alpha_s(l)$
when $l=\Lambda$. The terms in the sum~(\ref{PN-eq:selfenexp}) initially
decrease for $\alpha_s(m)$ small enough, but at some value of $n$ they start growing
again. The value at which the minimum is attained is easily obtained by use of the
Stirling approximation $n!\approx\sqrt{2\pi n} \exp(n\ln n -n)$.
The term of order $n+1$ in Eq.~(\ref{PN-eq:selfenexp})
can be written as
\begin{equation}
\label{PN-eq:selfenexp1}
\delta m^{(n+1)} = N m \alpha_s(m) \sqrt{2\pi} \exp\left[\left(n+\frac{1}{2}\right) \ln n
  -n + n\ln(2b_0 \alpha_s(m))\right],
\end{equation}
that has a minimum when the derivative of the exponent with respect to $n$ vanishes
\begin{equation}
\label{PN-eq:nmineq}
\ln n + \frac{1}{2 n} + \ln(2b_0 \alpha_s(m))=0\,.
\end{equation}
Under this condition the minimal term can be written as
\begin{equation}
\delta m_{\rm min} = N m  \alpha_s(m)\sqrt{2\pi} \exp\left[\frac{1}{2} \ln n_{\rm min}
  -n_{\rm min} -\frac{1}{2}\right].
\end{equation}
Solving eq.~(\ref{PN-eq:nmineq}), we get the value of $n$ at the minimum
\begin{equation}
n_{\rm min} = \frac{1}{2b_0\alpha_s(m)}-\frac{1}{2}+{\cal O}(\alpha_s(m))
\end{equation}
so that
\begin{equation}\label{PN-eq:minterm}
  \delta m_{\rm min} \approx  N  m \sqrt\frac{\pi\alpha_s(m)}{b_0} \exp\frac{1}{2b_0\alpha_s}
  =  N  \sqrt\frac{\pi\alpha_s(m)}{b_0} \Lambda.
\end{equation}
A rough procedure to sum divergent series is to sum up the terms as long as they decrease,
stopping at the minimum,
that therefore gives a first indication of the uncertainty in the result.

An alternative way to deal with factorially divergent expansion is by use of the Borel
transform. Given the power series
\begin{equation}
  f(\alpha_s)=\sum_{n=0}^\infty c_n \alpha_s^{n+1},
\end{equation}
the corresponding Borel transform is defined as
\begin{equation}
  B[f](t)=\sum_{n=0}^\infty c_{n} \frac{t^{n}}{n!}
\end{equation}
and we have formally
\begin{equation}
  f(\alpha_s)=\int_0^\infty {\rm d}t\, e^{-t/\alpha_s}  B[f](t).
\end{equation}
In the case of eq.~(\ref{PN-eq:selfenexp}), where $c_{n}=N(2b_0)^n n!$, we get
\begin{equation}
  m_{\rm P}-m = N m \int_0^\infty {\rm d}t\,e^{-\frac{t}{\alpha_s(m)}}\sum_{n=0}^\infty (2b_0)^n t^n
  = N m \int_0^\infty {\rm d}t\, e^{-\frac{t}{\alpha_s(m)}}\frac{1}{1-2b_0 t}.
\end{equation}
The presence of the pole along the real axis is again a manifestation of the renormalon
problem.

The Borel procedure gives an alternative, more educated method for the summation
of a factorially divergent power expansion: one takes as result the principal value
of the integral, and as uncertainty something proportional to the (absolute value)
of the imaginary part that arises
when the contour of integration is distorted above or below
the singularity (at $t=1/(2b_0)$ in our case) in the complex plane.
In ref.~\cite{Beneke:1998ui} it is proposed to take the absolute value of the imaginary
part divided by $\pi$ as the one-sided ambiguity, that in our case leads to
\begin{equation}\label{PN-re:borelamb}
N m  \frac{1}{2b_0} e^{\frac{1}{2b_0 \alpha_s(m)}}=\frac{N}{2b_0} \Lambda.
\end{equation}
In the following we will call this method for estimating the renormalon ambiguity
the ``Im/Pi method''.
In ref.~\cite{Beneke:2016cbu} the Im/Pi method is adopted, with the motivation that it works well
in context where the renormalon effect can be related to some physical observable
\cite{Beneke:1998ui}, and one can check with data that it gives reliable results.

Notice that the form of the minimal term in eq.~(\ref{PN-eq:minterm}) differs
parametrically by the one in eq.~(\ref{PN-re:borelamb}), due to the presence
of the extra factor $\sqrt{\alpha_s(m)}$.
In fact, in order to use the minimal term to estimate the ambiguity
of the result we should also account for the fact that several
terms of similar size may lie
near the minimum. Starting again from eq.~(\ref{PN-eq:selfenexp}), we can expand the
exponent around the minimum to get
\begin{equation}
  \delta m^{n+1}\sim \delta m_{\rm min} \exp\left[\frac{(n-n_{\rm min})^2}{2 n_{\rm min}}\right] \sim
  \delta m_{\rm min} \left[1+\frac{(n-n_{\rm min})^2}{2 n_{\rm min}}\right]\;.
\end{equation}
It is clear now that the size of the region where the terms of the series have similar
size is proportional to $\sqrt{n_{\rm min}}$. For definiteness, let us define this
region by requiring
\begin{equation}\label{PN-eq:fdef}
\left[1+\frac{(n-n_{\rm min})^2}{2 n_{\rm min}}\right]< f,\quad \mbox{with}\;f>1.
\end{equation}
This leads to
\begin{equation}
|n-n_{\rm min}| < \sqrt{2 (f-1) n_{\rm min}}\,
\end{equation}
and by multiplying the size of this region by the minimal term we get.
\begin{equation}
  \sqrt{2 (f-1) n_{\rm min}} N  \sqrt{\frac{\pi\alpha_s(m)}{b_0}} \Lambda
    =  N  \frac{\sqrt{(f-1)\pi}}{b_0} \Lambda
\end{equation}
This has the same parametric form of the ambiguity determined with the Im/Pi method,
and is in fact identical to it if we choose $f=1+1/(4\pi)$.

In the context of this oversimplified illustration of the mass
renormalon problem, we
also introduce the concept of the so called \PNmcMSR{} mass~\cite{Hoang:2017suc}.
We define it as
\begin{equation}\label{PN-MSRfull1}
 m_{\rm MSr}(R)=m+ N \int_R^m {\rm d}l\; \alpha_s(l)\,.
\end{equation}
or, equivalently
\begin{equation}\label{PNmc:eq-mpole-mmsr}
m_{\rm P}=m_{\rm MSr}(R) + N \int_0^R {\rm d}l\; \alpha_s(l)\,.
\end{equation}
Comparing eq.~(\ref{PNmc:eq-mpole-mmsr}) to eq.~(\ref{PNmc-eq:poleMSsimple}) and (\ref{PN-eq:selfenexp}), we see
that the difference between the pole mass and the \PNmcMSR{} mass
is given by the same power expansion as the difference between the pole mass and the \PNmcMSbar{}
mass, with the only difference that $\alpha_s$ is evaluated at the scale $R$ rather than
the scale $m$.

The \PNmcMSR{} mass looks like a formal interpolation between the \PNmcMSbar{} mass $m$ (when $R=m$) and
the pole mass (when $R=0$). However, at the low end, when $\mu=l$ the integral does not
exist. For $\mu>l$, on the other hand, the whole series is convergent, and does not
manifest any factorial growth for large $n$. In fact, the coefficients grow factorially
as long as $n<\log(m/\mu)$. For larger $n$, the factorial growth is damped out.

Since $m_{\rm P}$ is $\mu$ independent, we can also derive from~(\ref{PNmc:eq-mpole-mmsr}) an evolution equation
\begin{equation}
\mu  \frac{{\rm d} m_{\rm MSr}(\mu)}{{\rm d} \mu} = - N \mu \alpha_s(\mu),
\end{equation}
that, unlike the typical renormalization group evolution equations, has also a linear
dependence upon the scale $\mu$. This allows for the solution
to have a well-defined perturbative expansion in terms of the coupling evaluated
at some fixed scale, relative to its value at $\mu=0$.
On the other hand, it is also associated with the factorial growth
of its coefficients. Notice also that if we use the notation $m_{\rm P}(m)$ to denote
the pole mass,
and $m_{\rm MSr}(m,R)$ the \PNmcMSR{} mass of a heavy quark with
\PNmcMSbar{} mass $m$, we have
\begin{equation}\label{PNmc:eq-Rdep-mdep}
 m_{\rm P}(m) - m_{\rm MSr}(m,R) = m_{\rm P}(R)-R\,.
\end{equation}
In fact, the left hand side is equal to
\begin{equation}
  N \int_0^R {\rm d}l\; \alpha_s(l)\,,
\end{equation}
according to equation~(\ref{PNmc:eq-mpole-mmsr}), and the right hand side is equal
to the same quantity according to eq.~(\ref{PNmc-eq:poleMSsimple}) with $m$ replaced by $R$,
and $m_P$ taken as the pole mass corresponding to a \PNmcMSbar{} mass equal to $R$.

As a last point, we can ask what happens if there are other heavy flavours (typically
bottom and charm) below the top mass. The answer is quite obvious: we
just replace $\alpha_s(l)$ in eq.~(\ref{PNmc-eq:poleMSsimple}) with
a variable flavour $\alpha_s^{\rm vf}(l)$ that solves the evolution equation
\begin{equation}
  \mu^2 \frac{\rm d}{{\rm d} \mu^2} \frac{1}{\alpha_s^{\rm vf}(\mu)}=-b_0^{\rm vf}(\mu),
\quad b_0^{\rm vf}(\mu)=\frac{11 C_a - 4 T_f [3+\theta(\mu-m_c)+\theta(\mu-m_b)]}{12\pi}.
\end{equation}
We can immediately anticipate that the change in the mass relation will not
be dramatic, since it is induced by a relatively small change in $\alpha_s$ in a
relatively small region of the integration domain. However, the renormalon
uncertainty will be determined by eq.~(\ref{PN-re:borelamb}) with $\Lambda$
corresponding to the 3-flavour $\Lambda$. This is easily understood, since the
factorial growth at large order is controlled by increasingly small momenta, and
thus cannot be sensitive to flavours with large mass.

\subsection{The full story}
The Pole Mass $m_P$ is given in terms of the \PNmcMSbar{} mass $m$ by an expansion of the form
\begin{equation}\label{PN-eq:renormeq}
  m_P=m(\mu_m)\left\{1 + \sum_{n=0}^\infty
  c_n(\mu,\mu_m,m(\mu_m)) \alpha_s^{n+1}(\mu)\right\},
\end{equation}
where $m(\mu_m)$ is the \PNmcMSbar{} mass evaluated at the scale $\mu_m$,
and $\mu$ is the renormalization scale. The coefficients have been evaluated
up to the fourth order in $\alpha_s$~\cite{Marquard:2015qpa,Marquard:2016dcn}.
For the moment we assume that we have only one massive flavour, and all other are massless.
Formula~(\ref{PN-eq:renormeq}) was originally presented
in the 6-flavour scheme, where also divergences arising from the heavy flavour loops
are subtracted according to the \PNmcMSbar{} prescription. The coupling
constant is in this case the $(n_l+1)$-flavours (where $n_l$ is the number of light
flavours) coupling constant. The same result can be
expressed in the so called decoupling scheme, which is the scheme where divergences
caused by the heavy flavour loops are subtracted at zero momentum~\cite{Collins:1978wz},
in which case the coupling constant is the $n_l$ flavours one.

In order to obtain the decoupling scheme formula, it is enough to express the $(n_l+1)$-flavours
strong coupling in terms
of the $n_l$ one, according to the standard matching formulae (see
the QCD review in ref.~\cite{Patrignani:2016xqp}) and expand to the relevant order.
Since the matching conditions are known at four loops~\cite{Schroder:2005hy,Chetyrkin:2005ia}),
the full four loop accuracy of eq.~(\ref{PN-eq:renormeq}) can be retained.
From now on, we assume that
formula~(\ref{PN-eq:renormeq}) is expressed in the decoupling scheme. We stress that
also in this case heavy fermion loops do enter in the calculation,
but they are renormalized by zero momentum subtraction.

The leading IR renormalon divergence implies the following large-$n$  
behaviour of the perturbative coefficients~\cite{Beneke:1994rs} and~\cite{Beneke:1998ui}\footnote{Here we follow
  the notation of ref.~\cite{Beneke:2016cbu}}
\begin{eqnarray}
\label{PN-eq:cnasymp}
c_{n}(\mu,\mu_m,m(\mu_m)) & \underset{n\to\infty}\longrightarrow & 
N \frac{\mu}{m(\mu_m)}\,c^{(\rm as)}_{n}\,, 
\end{eqnarray}
where
\begin{eqnarray}
\label{PN-eq:ctildenasymp}
c_{n+1}^{(\rm as)} &=& 
(2 b_0)^n\, \frac{\Gamma(n+1+b)}{\Gamma(1+b)} 
\left(1+\frac{s_1}{n+b}+\frac{s_2}{(n+b)(n+b-1)}
+\cdots \right).
\end{eqnarray}
The $b$ and  $s_i$ coefficients can be found in ref.~\cite{Beneke:2016cbu}.
The $s_i$ coefficients of the 
sub-leading ${\cal O}(1/n^i)$ behaviour can all be given in terms of the 
coefficients of the beta-function~\cite{Beneke:1994rs}.
We notice that $m(\mu_m)$ cancels when inserting~(\ref{PN-eq:cnasymp})
in~(\ref{PN-eq:renormeq}).

Although the proof of the form of the asymptotic expansion is
non-trivial, it is not difficult to understand its properties. First of all,
the leading asymptotic behaviour arises from the region of small momenta
running in the loops. Since we are using a decoupling scheme, it is then
natural that the heavy quark mass and the scale at which it is evaluated
drop out. So, the only scale that can
appear in the mass correction is $\mu$.
If we neglect for the moment the mass of the light flavours, the form
in eq.~(\ref{PN-eq:cnasymp}) is the only one allowed by dimensional analysis.
The form of eq.~(\ref{PN-eq:ctildenasymp}) then follows by imposing that
in the derivative of the mass correction with respect to $\mu$ the
factorially growing terms should cancel at high orders, consistently with the
fact that factorial growth should arise from the low momentum region.

The coefficient $N$ cannot be computed exactly from first principles at the moment.
In the large $n_l$ limit (both positive or negative) it assumes the form~\cite{Beneke:1994sw}
\begin{equation}\label{eq:Nlargenl}
\lim_{|n_l| \to \infty} N=\frac{C_f}{\pi} \times e^{\frac{5}{6}}\,,
\end{equation}
which equals  $0.97656$ for $N_c=3$ ($C_f=4/3$). 
In ref.~\cite{Nason:2016tiy} it was noticed that the third and fourth known terms of the mass
conversion formula~(\ref{PN-eq:renormeq})
already show the renormalon asymptotic behaviour, and this can be used
to infer higher order terms in the top pole mass relation.
This was used
in ref.~\cite{Beneke:2016cbu} to extract the value of $N$ by fitting the
third and fourth order coefficient of the exact calculation
and give an improved
determination of the mass relation including the resummation of the
asymptotic terms.\footnote{
  Yet again, the observation of ref.~\cite{Nason:2016tiy} was made much earlier in a bottom physics
  context, and fits to extract the value of $N$ were performed
  even before the fourth order term was known, see refs.~\cite{Pineda:2001zq,Ayala:2014yxa}.}
The mass relation was determined by using the Borel
prescription, illustrated earlier.
The asymptotic expansion was evaluated by taking the principal value
in the inverse Borel transform formula. Then, its first four terms
where subtracted and replaced with
the exact ones. The renormalon ambiguity, obtained
according to the Im/Pi prescription,
was determined to be 70~MeV, in the case in which the bottom and charm quark mass effects are
neglected.

The procedure is modified when the charm and bottom mass effects are
included. I will not describe it here in detail. I only wish to remark
that in this case the renormalon ambiguity is obtained from the asymptotic
expansion of the pole mass relation with three light flavours. In other words, the
leading renormalon ambiguity only depends upon the number of light flavours, and nothing else.
It is thus the same for the top, bottom and charm pole masses, and it was determined
to be 110~MeV.

Adopting  $\alpha_s(M_Z)=0.1181 \pm 0.0013$ and
$m_t=163.508$~GeV (that yields $\alpha_s(m_t)=0.108531$ for the 5-flavour coupling constant,
the results can be summarized by the formulae
\begin{eqnarray}
  &\mbox{$m_c$, $m_b$ massless:}\quad &  \frac{m_{t,{\rm P}}}{m_t}=1.06164^{+0.00086}_{-0.00089}
                \pm 0.00043 \label{PNmc-eq:pole_o_msb_mcmb0}\\
  &\mbox{$m_c=1.3$, $m_b=4.2$~GeV:}\quad & \frac{m_{t,{\rm P}}}{m_t}=1.06213^{+0.00088}_{-0.00096}
                \pm 0.00066, \label{PNmc-eq:pole_o_msb}
\end{eqnarray}
where the last error is the irreducible ambiguity in the Borel integral, obtained
with the Im/Pi prescription, and the upper and lower
error on the central value are the sum in quadrature of all (reducible) errors in
the procedure, that are largely dominated by the uncertainty in $\alpha_s$.

In ref.~\cite{Hoang:2017btd}, the same task was dealt with a rather different method.
Before discussing it, it is useful to compare the final result
with that of ref.~\cite{Beneke:2016cbu}. For a top \PNmcMSbar{} mass of $m_t=163\,$GeV
used in \cite{Hoang:2017btd}, formulae~(\ref{PNmc-eq:pole_o_msb_mcmb0})
and~(\ref{PNmc-eq:pole_o_msb}) yield $m_{t,{\rm P}}=173.047$ and $173.127$
respectively, to be compared with the $173.086$ and $173.165$ values reported
on the last column of table 5 in ref.~\cite{Hoang:2017btd}, in the row
corresponding to the $R$ value equal to $m_t$. The final value they quote at
the end of section 4.4 is 173.186~GeV, about 60~MeV larger than the one obtained
here using the result of ref.~\cite{Beneke:2016cbu}, and thus well within the uncertainty.
On the other hand, they quote a larger uncertainty, of 180~MeV when the bottom and
charm masses are neglected, and of 250 MeV when they are taken into consideration.
It is instructive to examine the source of the differences.
For simplicity, I will focus upon the case of massless bottom and charm quarks.

In ref.~\cite{Hoang:2017btd}, an extensive use is made of the \PNmcMSR{} mass. In the case
of massless bottom and charm, this is defined as
\begin{equation}\label{PN-eq:msrmasshoang}
  m_P=m_{\rm MSr}(R) + R \sum_{n=0}^\infty
  c'_n(\mu,R,R)\, \alpha_s^{n+1}(R),
\end{equation}
where $\alpha_s$ is defined in the 5-flavours scheme, and the coefficients $c'_n$
are {\rm almost} the same as those in eq.~(\ref{PN-eq:renormeq}), since they are
given by the same set of Feynmann diagrams except for those involving top loops.
Was it not for this difference, we would have
$m_{\rm MSr}(m_{\rm MSr})=m(m)$.
This is the first of a number of subtleties involved in the definition of the \PNmcMSR{}
mass in ref.~\cite{Hoang:2017btd}. More subtleties come into play when the effects
of the bottom and charm masses are included. For the present discussion, however,
these subtleties are irrelevant. The important point is that we can express
$m_{\rm MSr}(R)$, for $R$ near the top mass, in terms of the \PNmcMSbar{} mass $m$,
and, as shown in our
elementary example, the \PNmcMSR{} mass obeys an evolution equation free of
renormalons, so that
it can be computed at any scale $R$ below the top mass, at least for not too low
values of $R$. Thus, we have the freedom to evaluate the pole mass
in eq.~(\ref{PN-eq:msrmasshoang}) at a scale $R$ of our choice, with the first
term evaluated using the evolution equation for $m_{\rm MSr}(R)$, while the second term
is evaluated using some prescription to handle the factorial growth
of the coefficients, that is again given by eq.~(\ref{PN-eq:cnasymp}).

The method used in~\cite{Hoang:2017btd} to estimate the asymptotic sum is
as follows:
\begin{enumerate}
\item Define: $m_{\rm P}(n)$ to be given by eq.~(\ref{PN-eq:msrmasshoang}) truncated
  up to (and including) the $n^{\rm th}$ order, and define $\Delta(n)=m_{\rm P}(n)-m_{\rm P}(n-1)$.
\item Find the minimal term $\Delta(n_{\rm min})$ and the set
  $\{n\}_f=\{n:\Delta(n) \leq f\times \Delta(n_{\rm min})\}$. The factor $f$ is
  defined to be ``larger but close to unity'', and the value $f=5/4$ is adopted, without further
  justification, in ref.~\cite{Hoang:2017btd}.
\item The midpoint of the covered range of values in  $m_{\rm P}(n)$ for $n\in  \{n\}_f$
  is used as central value, and the error is taken as half the size of the covered range.
  Scale variation is applied to the results within the range and included in the error.
\end{enumerate}
First of all, we notice that this method resembles the one presented in the previous
section, eq.~(\ref{PN-eq:fdef}), except for point 3. In fact, rather than taking half the
size of the range of values covered by $m_{\rm P}(n)$, it seems more natural to take the
sum of all $\Delta(n)$ divided by two.\footnote{Following literally the procedure of
  item 3 one would get a null error if the set $\{n\}_f$ consisted of a single
  element.}
In the following we will adopt this amendment for item 3, and furthermore
we will not consider scale variation effects, that we will discuss further on,
and refer to this procedure
as ``method D'', where D stands for ``discrete''. According to the discussion following
eq.~(\ref{PN-re:borelamb}), the Im/Pi method should roughly correspond to
method D, with the choice $f=1+1/(4\pi)$.

\begin{figure}
  \centerline{\includegraphics[width=0.46\textwidth]{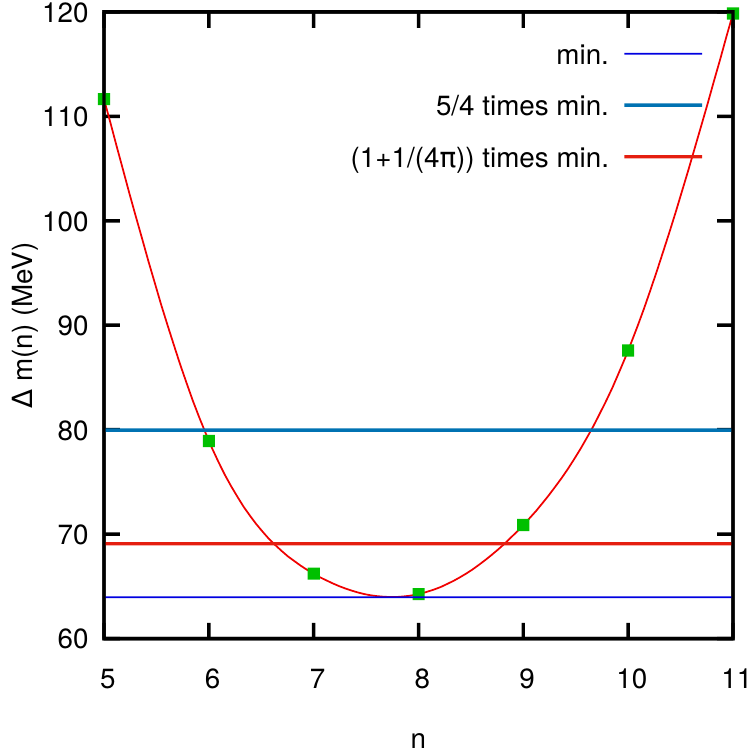}%
  \includegraphics[width=0.46\textwidth]{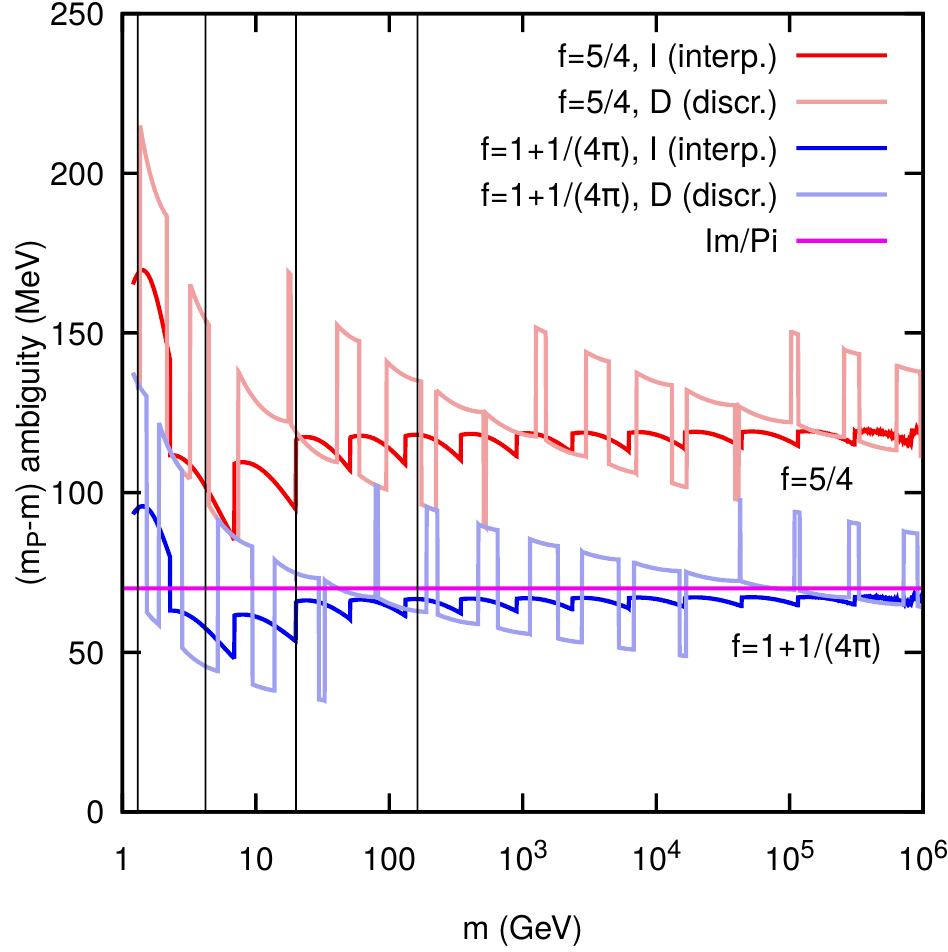}}
  \caption{In the left figure, the terms in the perturbative
    expansion for the pole mass relation are plotted as a function of the order.
    In the right figure, the pole mass ambiguity as a function of the pole mass,
    determined using the methods D and I described in the text are plotted
    as a function of the \PNmcMSbar{} mass $m$ for two choices of the
    parameter $f$.
    The value corresponding to the Im/Pi method is also shown.
    The vertical lines correspond to $m=1.3$, $4.2$, $20$ and $163$~GeV.}
  \label{PN-fig:Renorm}
\end{figure}

The situation is illustrated in fig~\ref{PN-fig:Renorm},
where the size of the terms for the perturbative expansion of the pole mass relation\footnote{%
  They are taken from ref.~\cite{Beneke:2016cbu}, table 2,
  multiplied by 163~GeV and by the factor $N=0.4616$, according to the same reference.}
are shown as a function of the order in the left plot.
In the right plot, we show the ambiguity in the top pole mass as a function of the top
mass $m$, taking $f=5/4$ or $f=1+1/(4\pi)$.
As suggested by eq.~(\ref{PNmc:eq-Rdep-mdep}), a plot analogous to the
right plot in fig.~\ref{PN-fig:Renorm} could be shown representing the
renormalon ambiguity, for a fixed \PNmcMSbar{} mass,
of the second term of eq.~(\ref{PN-eq:msrmasshoang})
as a function of $R$. Such plot is easy to obtain, and it is
indistinguishable from the one
shown in fig.~\ref{PN-fig:Renorm}. Thus I will not include it here.
The reader should thus keep in mind that the plot shown in fig.~\ref{PN-fig:Renorm}
is also representative of the renormalon ambiguity as determined in
ref.~\cite{Hoang:2017btd} as a function of $R$ instead of $m$, and that the vertical black lines in the
figure correspond to the specific values considered there.

The pole mass ambiguity should not depend upon $m$,
since it is determined by the small momentum region. However, because of the ``jumpy''
nature of the D prescription, we see considerable variations in the error
estimate, that get worse as the mass decreases. In order to remedy to this problem,
we can improve the procedure by extrapolating it to non-integer values of $n$.
This approach was also used in ref.~\cite{Beneke:2016cbu} in order to give an alternative
estimate of the sum of the asymptotic series, and turned out to give a result
that exceeded the one obtained with the Borel prescription by only 22~MeV.
Here I adopt a related procedure, that has the advantage of requiring a shorter and
more transparent explanation:
\begin{itemize}
\item
Calling $m_{\rm P}(n)$ the sum of the asymptotic expansion up to (and including) the $n^{\rm th}$ term,
one finds the value of $n_0$ such that $m_{\rm P}(n_0)-m_{\rm P}(n_0-1)$ is at a minimum. In other words,
the $n^{\rm th}$ term of the series is the smallest one.
\item
  One finds a cubic polynomial $P(n)$ such that $P(n)=m_{\rm P}(n)$ for $n=n_0-2$, $n_0-1$,
  $n_0$ and $n_0+1$. The polynomial $P(n)$ is taken as the extension of the series
  to non integer values of $n$. In particular, the value of the terms of the series
  at any $n$ are now given by ${\rm d}P(n)/{\rm d}n$.
\item
  One finds $n_{\rm min}$ such that ${\rm d^2}P(n)/{\rm d} n^2 =0$ for $n=n_{\rm min}$,
  and interprets it as the location of the minimal term.
  $P(n_{\rm min})$ is taken as the central value for the resummed result.
\item
  The error is taken equal to ${\rm d}P(n)/{\rm d} n$ evaluated at $n=n_{\rm min}$,
  times half the range
  in $n$ such that ${\rm d}P(n)/{\rm d} n$ does not exceed its minimal value
  by more than a factor $f$.
\end{itemize}
We call the above procedure ``method I'', where I stands for ``interpolation''.
The pole mass ambiguity obtained with this method is also shown in fig.~\ref{PN-fig:Renorm},
where it is seen to stabilize considerably the result down to masses of a few GeV.
It can also be noticed that, in an average sense,
it seems consistent with the discrete method.
At very low values of the mass, however, it also becomes unreliable. This is not unexpected: at
low value of the mass the minimal term occurs very early, when the series has
not reached its asymptotic form. Therefore, the use of a method based upon the size
of the minimal term cannot be recommended in this case.
On the other hand, the Im/Pi method always yields a constant ambiguity of $70$~MeV,
irrespective of the value of the mass. We notice that method I, with $f=1+1/(4\pi)$,
is fairly consistent with the Im/Pi method, as the simple argument following eq.~(\ref{PN-re:borelamb})
suggests.

A point that needs discussion is the method adopted in ref.~\cite{Hoang:2017btd}
to perform scale variation. The minimal term and the set $\{n\}_f$ is found
for the central value of the scale, and then the scale variation is performed on the result.
However, one should keep in mind that
also the set $\{n\}_f$ can be affected by scale variation, and its change should be included,
since it leads to scale compensation. This is an important point, that
deserves a more detailed discussion.

If we knew the whole perturbative expansion for the pole mass relation,
it would be formally independent of the renormalization scale.
The series, however, is only asymptotic, and we should
devise a method to sum it up. Such method better be scale independent too, so that the
result of the sum is also scale independent. It is easy to show with an example, borrowed
from our oversimplified model, that this is the
case for the Borel method. Let us assume that the top mass relation is exactly given by the formula
\begin{equation}
  m_{\rm P}=m+\int_0^m {\rm d} l\, \alpha_s(l),
\end{equation}
where $\alpha_s(l)$ is the one-loop strong coupling constant, and we assume that the only
massive flavour is the top. We know how to express the above mass relation in terms
of $\alpha(m)$, by using the Melling transform technique, that we now interpret as a formal,
order-by-order procedure:
\begin{equation}
  m_{\rm P}-m=\int_0^m {\rm d} l\, \alpha_s(l)=m\int_0^\infty  {\rm d}t \,e^{-t/\alpha_s(m)} \frac{1}{1-2b_0 t}.
\end{equation}
We now want to express the same relation using $\alpha(\mu)$, with $\mu$ of order $m$,
but different from it. This can be done as follows:
\begin{eqnarray}
  m_{\rm P}-m &=&\int_\mu^m {\rm d} l\, \alpha_s(l) +\int_0^\mu {\rm d} l\, \alpha_s(l) \nonumber \\
          &=&\int_\mu^m {\rm d} l\, \alpha_s(l) + \mu\int_0^\infty  {\rm d}t \,e^{-t/\alpha_s(\mu)} \frac{1}{1-2b_0 t}\,.
\end{eqnarray}
The first term can be expressed as a function of $\alpha_s(\mu)$ as long
as $\mu>\Lambda_{\rm QCD}$, and in fact it has a convergent expansion in terms
of $\alpha_s(\mu)$, while the second term has the divergent, asymptotic expansion,
still in terms of $\alpha_s(\mu)$.
We now show that the two expressions, if the Borel integral is evaluated according to
the principal value prescription, are identical, i.e. the scale variation of the
series resummed according to the principal value in the Borel integral is zero.
In order to prove it, we take the difference of the two formulae
both regulated according to the principal value prescription.
We get
\begin{equation}
  \int_\mu^m {\rm d} l \alpha_s(l) +\int_0^\infty  {\rm d}t \,\left[\mu e^{-t/\alpha_s(\mu)}-m e^{-t/\alpha_s(m)}\right]
  \frac{1}{1-2b_0 t} = 0\,.
\end{equation}
But now we see that the principal value prescription is no longer needed,
since both terms in the square bracket
become equal to $\Lambda_{\rm QCD}$ for $1-2b_0 t=0$. At this point we have to show that the expression vanishes.
It certainly vanishes for $\mu=m$, and its derivative
with respect to $\mu$, given by
\begin{eqnarray}
&&  -\alpha_s(\mu) + \int_0^\infty  {\rm d}t \,\left[e^{-t/\alpha_s(\mu)}\left(1+\mu\frac{\partial}{\partial \mu}\frac{-t}{\alpha(\mu)}
    \right)\right]\frac{1}{1-2b_0 t} \nonumber \\
  &=& -\alpha_s(\mu) + \int_0^\infty  {\rm d}t \,\left[e^{-t/\alpha_s(\mu)}\left(1-2b_0t\right)\right]\frac{1}{1-2b_0 t}
 \nonumber \\  &=& -\alpha_s(\mu) + \int_0^\infty  {\rm d}t e^{-t/\alpha_s(\mu)}\,=\,0,
\end{eqnarray}
vanishes also, and thus we have full scale independence of the result.

\begin{figure}
  \centerline{\includegraphics[width=0.48\textwidth,page=1]{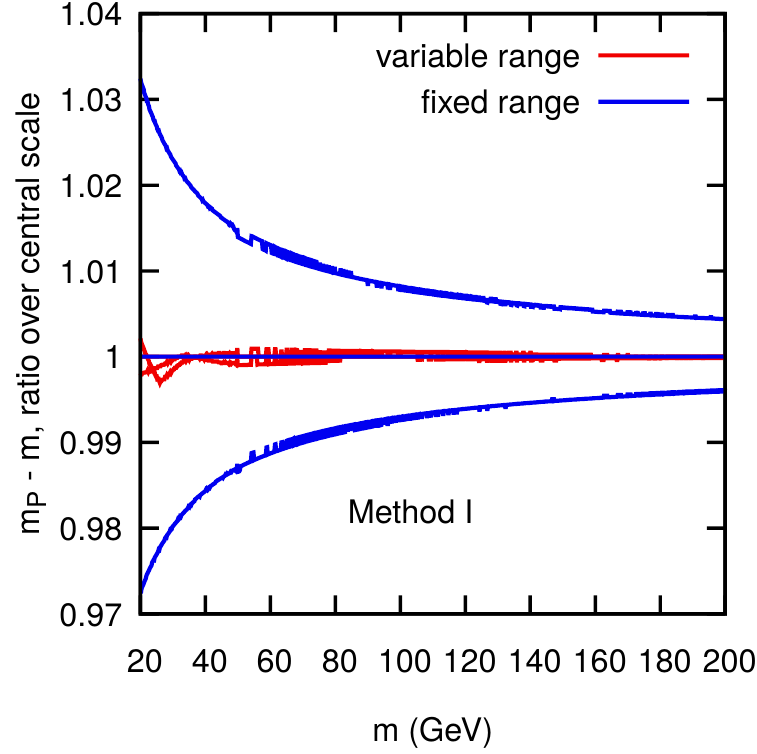}%
  \includegraphics[width=0.48\textwidth,page=2]{Figs/plotscalevar}}
  \caption{Scale dependence of the pole - \PNmcMSbar{} mass difference.
    The scale is varied by a factor of two above and below $m$.
    The curves labelled as ``fixed range'' are obtained by holding the
    position of the minimal term and of the range fixed while varying the
    scale; the curves labelled as ``variable range'' are obtained by recomputing
    the minimal term and the range for each scale choice.
    In the case of method $I$ the result does not depend upon $f$,
    while in the case of method $D$ a mild dependence upon $f$ is found,
    as can be easily understood from the definition of the method.}
  \label{PN-fig:scalevar}
\end{figure}

In resummation methods that rely upon the minimal term, scale independence is no longer
exact, since a truncation of the series is involved. But it is quite clear that the location
of the minimal term must be kept variable with the scale, since it leads to scale compensation.
This is illustrated in fig.~\ref{PN-fig:scalevar}.
The figure was obtained from eq.~(\ref{PNmc-eq:poleMSsimple}), and its
expansion given by the terms in eq.~(\ref{PN-eq:selfenexp}), by
expressing $\alpha_s(m)$ as a function of $\alpha_s(\mu)$ (using the
leading order formula) and expanding again in powers of
$\alpha_s(\mu)$, a procedure that can be easily performed using
algebraic manipulation programs. The scale $\mu$ is taken equal
to $2m$ and $m$.
The series can then be resummed
using our prescriptions D and I that rely upon the minimal term. We
consider the two approaches: one where we
recompute the location of the minimal term and of the range
for each scale choice, and one where we do not,
this last one corresponding to ref.~\cite{Hoang:2017btd}.
It is clear that in both the D and I methods, further scale compensation takes place
if we recompute the location of the minimal term and the range when
we vary the scale. When using method I, the scale variation
almost disappears, which is consistent with the observation that this method
is fairly close to the Im/Pi prescription. Again, in the D method, due to its
jumpy nature, scale
compensation, although considerable, is less effective.
When going at very low value of the mass, both
methods become quite unstable, again suggesting that it is better to avoid
the use of minimal term based prescriptions in this case.

In summary, although some freedom in the determination of the
renormalon ambiguity cannot be avoided, one can identify
three aspects in the procedure of ref.~\cite{Hoang:2017btd}
that should be amended:
the choice of the
parameter $f$; the insistence in evaluating the sum of the asymptotic expansion
at low scales using a minimal term based, discrete method;
and  the procedure adopted to determine scale variation
uncertainty.
As far as the choice of $f$ is concerned, some justification for the
chosen value is clearly missing. As far as going to low scales for determining the
renormalon ambiguity, it is clear that when using resummation methods based upon
the minimal term there is essentially no limit to the size of the ambiguity
that one can get, since at sufficiently low scales the first term
of the expansion is the minimal term, and it can become as large as one pleases
because of the Landau pole.
Finally, the method adopted for the scale variation uncertainty does not
include effects (i.e. the displacement of the minimal point and the range)
that lead to scale compensation.

As a conclusion of this discussion, we may wonder whether the issue of the size
of the mass renormalon ambiguity will ever have any importance at the LHC.
If we are arguing about whether the ambiguity is 110 or 250 MeV, this seems,
at the moment, very unlikely. While the mass renormalon is now the only non-perturbative
ambiguity that we can discuss to such detail, there are
certainly other non-perturbative effects, very likely associated with renormalon
uncertainty, that are much less understood, and that pose a much more serious threat
to the accuracy of the top mass determination at the LHC.

\section{Conclusions}
The problem of the top mass measurement at hadron colliders is
certainly a very subtle one, and has received the attention of many
researchers.  At the moment, there is no wide consensus in the
theoretical community on how it should be dealt with.

In the present work I have reviewed the current theoretical status,
highlighting several issues that need to be resolved. In the following,
I summarize my view on the subject.

In Monte Carlo generators the
top decay products are distributed with a Breit Wigner distribution.
Thus, the corresponding mass parameter should be qualified as the top
pole mass. It is clear that Monte Carlos represent physics processes only in an
approximate sense, and thus, if we use them to extract the pole mass by fitting mass
sensitive distributions, the result will
be subject to errors due to Monte Carlo inaccuracies.
However, this is true of \emph{any} calculation of
a physics process. Thus, I do not accept the view that, because of
these inaccuracies, one should qualify the fitted mass parameter as a ``Monte Carlo
mass'' rather than the top pole mass.  What one should instead do is
to ask how the approximations and inaccuracies of the Monte Carlo
propagate into mass sensitive observables, thus leading to an error on
the extracted top pole mass. Observe that this is the same approach
that we adopt when we perform fixed order calculations, or resummed
calculations at various level of precision.
There is no reason why Monte Carlo generators should be
treated differently.

We can loosely separate the approximations and inaccuracies of the Monte Carlo
into two categories: those that affect the perturbative part, and those that affect
hadronization phenomena and their matching to the perturbative part.

In relation to the Monte Carlo inaccuracies and approximations in the
perturbative part, much has been done in order to understand their
impact. We have now at our disposal generators of increasing
perturbative accuracy, that include only the shower, the
shower with the inclusion of matrix element corrections (MEC), or
the shower matched to an NLO calculation (NLO+PS)
of top production and decay.  By comparing
them, we can assess the errors associated with the less accurate
generators, and at the same time get an estimate of the remaining
inaccuracies.

It is clear that for the standard measurements of the
top mass, the most relevant component of the perturbative model is the
formation of the $b$ jet.  Thus, a shower Monte Carlo that handles
production with only leading order accuracy, but performs matrix
element corrections to top decay, may be quite adequate for this
purpose. Thus, it is important to examine the
Monte Carlo accuracy for the observable at hand.  By relying upon the
generic statement that Monte Carlos have only leading order accuracy,
we simply miss the point.

Notice that in higher orders calculations to be used for top mass
extraction in standard measurements, such as those that enter the
NLO+PS generators, we \emph{must} use a mass scheme such that the top
mass is close enough to the
pole mass (i.e. that differs from it by less than the top width),
otherwise an important shift in the mass of the
reconstructed top\footnote{Here, by ``reconstructed top'' I mean a
  system defined at the particle level that is likely to have
  originated from the decay of a top quark.  We may define it, for
  example, as the hardest $b$-jet, the hardest positive lepton and the
  hardest neutrino of matching flavour.}  at the next perturbative
order would systematically arise.  Thus, we might as well use the pole
mass scheme in these contexts.

More subtle effects, not correctly modelled with the Monte Carlo, have
to do with interference of radiation in production and decay. It is
unclear in this case whether the radiated parton should be considered
part of the system of the top decay products (this phenomenon is a
perturbative precursor of the non-perturbative color reconnection
problem).  The worry is that these emissions, if not implemented
correctly, may induce systematic shift in the mass of the
reconstructed top that would be difficult to characterize.

In recent
times, techniques to deal with interference effects in production and
decay within the framework of NLO+PS generators have become
available~\cite{Jezo:2015aia,Frederix:2016rdc}, and have been used to
build a $t\bar{t}$ generator~\cite{Jezo:2016ujg}.  This has been
compared to a generator that does not include these
features~\cite{Campbell:2014kua}, and it was found
to yield very similar results. This can be taken as evidence
that interference in production and decay has little impact on top mass
measurements.

In a recent study~\cite{Ravasio:2018lzi}
it has been shown that the
new generator of ref.~\cite{Jezo:2016ujg}, interfaced with Pythia8, yields results
on the mass of the reconstructed top
that differ from the ones obtained with {\tt POWHEG-hvq} interfaced with Pythia8 (that is now
the default of the LHC experimental collaborations) by less than 50~MeV,
if one assumes to have full access to the ``particle truth'' level,
or less than 200~MeV, if the experimental resolution is taken into
account. Results like this give us confidence that we have control
over the perturbative side of our generators. They should
be challenged and scrutinized by the theoretical community, also by building
and trying different generators with the same or better accuracy.

Monte Carlo inaccuracies due to the modeling of non-perturbative
effects, and to their interplay with renormalons, are certainly harder to examine.
Even there, however, some recent progress has taken place. The calculation of the
fourth order term in the relation of the pole mass to the \PNmcMSbar{} mass,
yielding a contribution of 200~MeV, has allowed to make reliable projections
on the size of higher order terms, to the point where one can be confident that
the pole mass can be safely used for the LHC top mass measurements.
I recognize, however, that this is just one piece of the puzzle.
Other power suppressed effects may be more tricky to estimate.

Notice that estimating a ``perturbative precursor'' of a
non-perturbative effect does not necessarily yield an upper bound on
the latter.  For example, the fact that at leading
order interference between radiation in production and decays is
small does not imply that colour reconnection effects are
small. This counter-intuitive fact may be also understood if we assume
that non-perturbative effects are associated to renormalons, in which
case a contribution at a given
perturbative order may be followed by others that decrease less than
geometrically and have all the same sign. Under these conditions, the given
contribution cannot be considered a good estimate of the error due to
the missing terms of higher order.

It is clear that more theoretical work is badly needed in this
framework, with new ideas on how to estimate linear power suppressed
effects and their interplay with renormalons. The works of
ref.~\cite{Hoang:2008xm} and its followups go in the right direction,
but should be extended to cover top quarks of moderate transverse momenta
in order to be useful for standard top mass measurements, and should
focus upon assessing the errors associated with direct measurement
techniques.

There are other kinds of studies that are, in my opinion, even more
urgent. Since Monte Carlos do fit the data, and since there is much
arbitrarity in the way they model non-perturbative effects, studies
that exploit this arbitrarity to estimate the associated
uncertainties are badly needed. The inclusion of alternative models of
colour reconnection in Pythia8 is one such example. These alternative
models have been used by CMS, and have led to an increase of the
systematic error of the top mass determination in the recent 13~TeV
measurement. Needless to say, these model should undergo further
scrutiny by the theoretical community. We should examine and criticize
them, and reach a consensus on whether they are acceptable, too
extreme or too narrow.

I firmly believe that more studies of this sort are needed. They should proceed as follows:
\begin{enumerate}
\item Choose a parameter that is a matter of concern for the top mass measurement.
  Rather than varying a parameter, one may also consider more drastic variations,
  like changing a full component or the whole generator with a different one.
\item Vary the parameter and re-tune the generator (or change the generator).
  Restrict the range of variation of the parameter so that an acceptable tune can be achieved
  (or make sure that the new generator gives a good description of the data).
\item Set up a simplified framework to assess the impact on top mass measurement,
  and determine how much the extracted mass changes with the new setting.
\end{enumerate}
Notice that this implies a lot of work, and furthermore,
by proceding in this way we may end up uncovering problems
that we have not yet foreseen. In other words, it takes courage to do it.
However, I believe that this is the only way to set up a reliable
method to assess the uncertainties, and gain an increased
confidence on the reliability of our result.\footnote{%
  This can also guide us in improving the measurement method,
  along lines similar to those explored in ref.~\cite{Andreassen:2017ugs}.}

As an example, in the work of ref.~\cite{Ravasio:2018lzi} it was found that
by using Herwig7 rather than Pythia8 as shower generator, the
reconstructed top peak according to the particle level truth (i.e.
assuming that we can detect and identify all final state particles including
neutrinos and $b$ hadrons) is displaced by less than 100~MeV when using the new generators
of ref.~\cite{Jezo:2016ujg} and \cite{Campbell:2014kua}, and by less than 250~MeV
when using the old {\tt POWHEG-hvq} one. This result is quite remarkable, since the
shower models (and, as a consequence, their interface to hadronization) and the
hadronization models themselves are totally different in the two generators, and yet they
cause a displacement in the top reconstructed peak that is significantly below the 1~GeV
level. The same
study also finds important differences, of the order of 1~GeV, if experimental resolution
effects are taken into account. This should be considered, however, a less severe problem,
since in principle this error could be reduced by increasing the resolution.~\footnote{%
  Furthermore, in the same study some evidence was found that it may not be possible
  to fit the same data set with both generators. In other words, it may not be
  possible to satisfy item (2) in the above list with both generators.}

To conclude, let me say again that
I am very aware that this is a controversial subject. The overall view
that I have expressed should be taken as my personal one. However, several
points that I made are shared by many theorist colleagues, and my current opinions
have been deeply influenced by discussing with them. None of these views have
been expressed in writing so far, so I have just done it here, in the
hope that a more transparent debate on this subject will be started,
and some wider consensus may be reached in the future.

\section{A personal note}
My last exchange with Guido was by e-mail, on the 16th of April
2015, when I learned that he had been awarded the EPS High Energy and
Particle Physics Prize. I did not know of his disease at that
time. I sent him a very synthetic e-mail, that had a 'CONGRATULAZIONI PER L'EPS!' in
the subject, and ``Era ora ...  ciao, Paolo'' in the body. His reply
was particularly warm: ``Caro Paolo, ti ringrazio moltissimo. Sei un
caro amico. Saluti G''. I now find this exchange particularly moving.
\vskip 0.3cm
\noindent{\bf Acknowledgements}
\vskip 0.1cm
I wish to thank Giulia Zanderighi, Gavin Salam, Matthias Steinhauser
and Michelangelo Mangano for reading the manuscript, and for providing
useful suggestions.

\providecommand{\href}[2]{#2}\begingroup\raggedright\endgroup
\end{document}